\documentclass[11pt,twoside]{article}

\usepackage{asp2006}
\usepackage{epsf}
\usepackage{lscape}

\begin{document}
\title{Theoretical problems and perspectives}   %%% Fill in title
\author{Francoise Combes}   %%% Fill in author names
\affil{Observatoire de Paris, LERMA, 61 Av. de l'Observatoire, F-75014 Paris, France}    %%% author affiliations

\begin{abstract} %%% 
This talk tries to summarise where we are now, in the ``nature and nurture'' questions in galaxy
formation and evolution, and briefly describe unsolved problems, and perspectives of progress.
\end{abstract}

%%% MAIN BODY OF TEXT GOES HERE. CONSULT "INSTRUCTIONS FOR AUTHORS USING

%\section{}   %%% Top level section head (remove "%" symbol)
%\subsection{}   %%% Second level section head (remove "%" symbol)
%\subsubsection{}   %%% Lowest level section head (remove "%" symbol)
%\section*{}    %%% Unnumbered top level section head (remove "%" symbol)
%\subsection*{}   %%% Unnumbered second level section head (remove "%" symbol)

\section{Possibly solved questions}

Throughout this meeting, there has been a lively debate about the mere definition 
of what is an isolated galaxy, and it appears that a consensus has
arisen that  although complete isolation does not exist, at 
least the most isolated objects can be identified.
So the questions where an answer has been proposed in this conference 
are the following:

\begin{itemize}
\item Are there isolated galaxies in the universe? 
Robust definitions and criteria, involving relative separation of galaxies,
in both space and velocities, have been proposed, ensuring
isolation at least for time lapse of billion years (talk by Karachentseva, AMIGA contributions).
\item Is there a void problem, i.e. do we see galaxies in the voids, as 
predicted by the standard $\Lambda$CDM theory and simulations (e.g. Peebles, 2001)?  We have 
heard in several talks (Tinker, Croton)  that the problem is solved,  at least at zeroth order:
environment is a secondary parameter in simulations. However, due to the morphology
segregation and bias, we should see some dwarf galaxies in the voids, that  
 are not there (Koribalski).
\item The existence and life-time of compact groups (CG) have always been
an issue; but  genuine CG are an ideal location to study environmental effects.
May be two thirds of apparent CG are only chance alignment (and one third only
if selected in velocity, Mamon), the special dense environment of CG 
is visible in peculiar colors, star formation, morphology (McConnachie),
 and low fraction of broad line AGN (Dultzin, Martinez).
\item How are formed isolated early-type galaxies (ETG)? are they the end evolution of
fossil groups?  20\% ETG lie in low density environment, some with X-ray haloes,
some have recently assembled (Forbes); their evolution might depend more on 
halo mass than environment.
\end{itemize}

\section{Remaining questions}

 Several issues remain to be solved in the standard 
galaxy formation scenario, which are intimately related
to the environment. Let us mention successively how the
 luminosity function varies with environment, and whether
the over- prediction of bright and faint galaxies could be solved,
how the bimodality between red and blue galaxies,
and in particular the mass limit separating the two varies
 with environment,
how the galaxy density, through harassment and strangulation can
produce star formation quenching, leading to downsizing. 
The problem of bulge-less galaxies depends strongly on environment,
very large fraction of them are found in isolated galaxies.

\subsection{Mass and light distribution function}

One of the main problems encountered in
the $\Lambda$CDM scenario (numerical and semi-analytical simulations)
is the prediction of too many bright and too many faint galaxies
(Baugh 2006, Eke et al 2006, Jenkins et al 2001).
To avoid the formation of stars in dwarf haloes, models rely on star
formation feedback.
Gas is heated in dwarfs, but then falls in heavier haloes,
which worsens the bright end problem.
To limit the star formation and baryon infall in
massive galaxies, requires AGN feedback 
(e.g. Somerville et al 2008).
Croton \& Farrar (2008) show how the
luminosity function varies with environment:
at first order, the profile with respect to mass
is the same for high and low density, but all populations
are 10 times lower in voids. The function is dominated by
the blue late-type galaxies at the faint end. Due to the
shift to low-mass galaxies in voids, the blue ones are then
predominant.

The VIMOS deep survey (Ilbert et al 2006) reveals that
at about half-life of the universe, the morphological
segregation of galaxies according to over or under-dense regions
is already pronounced. Nurture effects should act quite early
or nature is important.

\subsection{Bimodality}

The distinction between red sequence and blue cloud
galaxies in the SDSS (Baldry et al 2004) covers
many parameters, essentially  the star formation rate
and history, but also dust, age, metallicity.
The two classes are explained by
two different formation mechanisms,
with a separating stellar mass limit of 3 10$^{10}$M$_\odot$.
The fraction in red sequence increases both with mass and environment
(Baldry et al 2006).

An alternative interpretation is that
star formation history depends on surface density
(Kauffmann et al 2003).
 There is a transition in disk surface density (300 M$_\odot$/pc$^2$), where the gas
begins to outflow, corresponding to the supernova escape velocity 
of about 100km/s.

The origin of the bimodality has been proposed in terms
of gas accretion from external filaments:
above a certain mass (3 10$^{11}$ M$_\odot$ in dark matter halo), the gas is 
not accreted cold, but is heated in shocks and  
has no time to cool, while below cold gas is accreted
(Keres et al 2005, Dekel \& Birnboim 2006). However,
the bimodality  is not well reproduced by semi-analytical models:
there is an excess of blue bright objects, and red faint satellites
(de Lucia al 2006).

\subsection{Downsizing}

The well-known paradox for hierarchical scenarios, i.e.
that massive galaxies  have  the shortest formation timescales, 
might be interpreted in terms
of environment. Comparing the star formation history of field
and cluster ellipticals,  the formation time is delayed in the field
(de Lucia et al 2006).
The hierarchical formation of the brightest galaxy in clusters
have been followed by de Lucia \& Blaizot (2007): most stars
are formed before z=5, but the mass is mainly assembled after
z=0.5 through dry mergers. 

In a large SDSS sample, stellar ages and metallicities have been studied,
with respect to mass and environment (Mateus et al 2007).
In clusters, massive early-type galaxies are older and more metallic,
they have formed at high redshifts;
galaxy evolution is accelerated in denser environments, and
proceeds via a nurture way.

\subsection{Problem of bulgeless galaxies}

One of the hard problems encountered by the hierarchical scenario
is the high frequency of bulge-less spiral galaxies today. Those must not have  
experienced major mergers.
Locally, about two thirds  of the bright spirals are bulgeless, or have a light bulge
(Kormendy \& Fisher 2008, Weinzirl et al 2009).
Surprising is the large frequency of edge-on superthin galaxies (Kautsch et al 2006).
one third  of galaxies are completely bulgeless.
Among the SDSS sample, 20\% of bright spirals are bulgeless until z=0.03 
(Barazza et al 2008). The frequency is of course higher 
in low-density environment (talk by Karachentsev, this meeting).

These numbers are not compatible with the predicted bulge-to-total mass
ratio predicted in semi-analytic models, with major mergers (Weinzirl et al 2009).
In $\Lambda$CDM, a bulge-to-total mass ratio lower than 0.2 requires 
no merger since 10 Gyr (or last merger before z=2). The
 predicted frequency of bulge-less galaxies is 15 times lower  than observed. 

The solution might reside in the importance of
cold gas accretion on galaxy mass assembly (Keres et al 2005).
The relative fraction of mass accreted from cold gas in filaments
is larger in low-mass haloes, and therefore in under-dense regions.
Both observations and simulations reveal that
most of the  starbursts are due to smooth flows (e.g. Robaina et al 2009).
Inflow rates are sufficient to assemble galaxy mass
(10-100 M$_\odot$/yr).
 Gas accretion can occur at arbitrary angles, 
and form sometimes polar rings. Stanonik et al (2009) show
such a  gaseous polar disk, around a
galaxy aligned along a wall between voids
(cf van de Weygaert, this meeting  and the winning poster!).
Gas from the cosmic filaments is flowing to the wall,
perpendicular to it.

\section{Perspectives}

We are now confident that the most isolated galaxies
can and have been identified, and the respective roles of nature and nurture
have been recently well highlighted.

\begin{itemize}

\item Nature is revealed by
faster evolution and merging in over-dense regions, that will become clusters,
while the nurture effects are obvious through strangulation, ram-pressure, harassment,
once clusters are formed.

\item Star formation and AGN feedback are necessary 
to fit the luminosity function of galaxies, and baryon fraction  in stars.

\item The downsizing paradox is partly due to environment, but semi-analytic 
models have still too many bright blue objects at z=0, and too many red faint satellites.
May be gas accretion should not be stopped for these faint satellites,
which would enhance also the green valley.

\item The role of mergers might have been over-estimated with respect
to gas accretion in galaxy mass assembly, and this could explain the 
bulge-less galaxy problem.

\end{itemize}

\acknowledgements %%% Text of acknowledgements runs on after this command.
Many thanks to the organisers for the invitation to such a lively and friendly meeting.

\end{document}